\documentclass[preprint,12pt]{elsarticle}
\usepackage{amssymb}
\usepackage{epsfig}
\usepackage{amsthm}
\usepackage{amsmath}
\journal{Physica A}
\begin{document}
\begin{frontmatter}

\title{Diffusion on a flat potential with a new localized sink: Exact Analytical Solution}

\author{Hemani Chhabra \corref{cor1}}
\address{Theoretical Chemistry Group, 
University of Oxford, Oxford, UK,}

\author{Aniruddha Chakraborty\corref{cor2}}
\address{Indian Institute of Technology Mandi, Kamand, Himachal Pradesh - 175005, India.}
\cortext[cor1]{Corresponding author}

\fntext[fn1]{Phone:-+91-9643447789}

\begin{abstract} 
We give a method for finding the exact analytical solution for the problem of a particle undergoing diffusive motion in a flat potential in the presence of a new localized sink. The Diffusive motion is described using the Smoluchowski equation. Our method requires the knowledge of Laplace transform of Green's function for the motion in absence of the sink. The same model for sink can be used to deal with other potentials.
\end{abstract}

\begin{keyword}flat potential; localized sink; analytical; Greens function.
\end{keyword}\end{frontmatter}

\section{Introduction}
\noindent Understanding reaction-diffusion system holds significant importance in almost all areas of science.  The most simplest model to study reaction-diffusion system is to use Smoluchowski equation for an appropriate potential in the presence of a sink \cite{Risken}.
It serves as a model to study various chemical dynamic \cite{Nishijima,Oster} processes in solution. There are only few potentials for which the exact analytical solution of Smoluchowski equation (without any sink) is known. For those cases, various methods have been proposed to solve analytically the Smoluchowski equation with different types of sinks e.g., pinhole, gaussian \cite{Bagchi2} and lorentzian sink \cite{Bagchi} and Dirac delta function sink \cite{Sebastian1,Sebastian2,Chakravarti}. In the following we propose a new localized sink, and find the exact analytical solution in the case of a flat potential. In different situations different type of sinks are the appropriate one for describing that particular reaction diffusion system. It is very difficult to get an analytical expression for survival probability for any model, in the following we are proposing a new localized sink for which exact analytical expression for survival probability is derived in this paper.

\section{Smoluchowski equation for a flat potential with sink}

\noindent We would like to solve the following Smoluchowski equation

\begin{equation}
\frac{\partial P(x,t)}{\partial t} = (L-K_0S(x)-K_r)  P(x,t),
\end{equation}
where 
\begin{equation}
L=D \frac{\partial^2}{\partial x^2} + \frac{D}{k_b T} \frac{\partial}{\partial x} \left({\frac{\partial V(x)}{\partial x}}\right).
\end{equation}
Here $P(x,t)$ is the probability that the particle may be found at the position $x$ at time $t$, $V(x)$ is the potential responsible for the motion of particle. $S(x)$ is the position dependent sink function, assuming it to be normalized. $K_0$ and $K_r$ are the rates of non-radiative decay and radiative decay respectively.
$D$ is the diffusion coefficient which is derived in relation to the friction coefficient($\zeta$), where $D=\frac{k_bT}{\zeta}$ and $T$ is the temperature \cite{Sebastian2}. Now we do Laplace transformation of $P(x,t)$ using 
\begin{equation}
\tilde P(x,s)= \int^\infty_0 P(x,t) e^{-st} dt.
\end{equation}
Laplace Transformation of Eq. (1) yields the following equation
\begin{equation}
P(x,0)=(s-L+K_0S(x)+K_r)\tilde P(x,s).
\end{equation}
\noindent Now for a flat potential 
\begin{equation}
\frac{\partial V(x)}{\partial x}=0.
\end{equation}
And assuming only non-radiative decay taking place, therefore $K_r=0$. The effect of this non-radiative decay term can be incorporated in our calculation very easily, wherever required. Then on rearrangement and applying above mentioned conditions Eq. (4) reduces to
\begin{equation}
s\tilde P(x,s)-P(x,0)=D \frac{\partial^2\tilde P(x,s)}{\partial x^2} - K_0 S(x)\tilde P(x,s).
\end{equation}
Assuming P(x,0) to be a Dirac delta function. Therefore, Eq. (6) is now modified as
\begin{equation}
s\tilde P(x,s)-\delta(x+a)=D \frac{\partial^2\tilde P(x,s)}{\partial x^2} - K_0 S(x)\tilde P(x,s).
\end{equation}
In the following we will solve Eq. (7) in case of localized sink.

\section{Exact solution for localized sink }
\noindent In the following we assume that S(x) is non-zero for a short range of x-values around x=0,
\begin{equation}
S(x)=S(0)+\left(\frac{d S}{d x}\right)_{x=0}+......
\end{equation} and similarly
Ignoring all higher order terms (other than zero order term) in the Taylor series expansion of $S(x)$ and replacing the terms $S(x) \tilde P (x,s))$ by $S(0) \tilde P (0,s))$, Eq. (7) is now modified as 
\begin{equation}
s\tilde P(x,s)-\delta(x+a)=D \frac{\partial^2\tilde P(x,s)}{\partial x^2} - K_0 S(0)\tilde P(0,s) f(x).
\end{equation} 
Where $f(x)$ equals to $1$ for $x$ values between $0-\epsilon$ and $0+\epsilon$ and $f(x)$ equals to zero otherwise, where $\epsilon$ is a small positive number. The Eq. (9) is now solved using the Green's function method.  
\begin{equation}
\tilde P(x,s)= \int^\infty_{-\infty} dx_{0}G(x,x_0|s)\delta(x_0+a) - K_0S(0)\tilde P(0,s)\int^{\infty}_{-\infty} dx_{0}G(x,x_0|s)f(x_0).
\end{equation}
Using the analytical expression of Green's function \cite{Diwaker2} the above equation can be simplified as
\begin{equation}
\tilde P(x,s)=\frac{e^{-\sqrt{\frac{s}{D}}|x+a|}}{2\sqrt{sD}} -\frac{K_0S(0)\tilde P(0,s)}{2\sqrt{sD}}\int^{0+\epsilon}_{0-\epsilon} e^{-{\sqrt{\frac{s}{D}}|x-x_0|}}dx_0.
\end{equation}
Which on integrating over $x_0$ (with the assumption that $\epsilon$ is a very small positive number)
\begin{equation}
\tilde P(x,s)=\frac{e^{-\sqrt{\frac{s}{D}}|x+a|}}{2\sqrt{sD}} -\frac{K_0S(0) \epsilon}{\sqrt{s D}}e^{-\sqrt{\frac{s}{D}} |x|}\tilde P(0,s).
\end{equation}
For $x=0$ the above equation becomes
\begin{equation}
\tilde P(0,s)=\frac{e^{-\sqrt{\frac{s}{D}}|a|}}{2\sqrt{sD}} -\frac{K_0S(0) \epsilon}{\sqrt{s D}}\tilde P(0,s).
\end{equation}
We now solve the above equation for $\tilde P(0,s)$ to get  
\begin{equation}
\tilde P(0,s)=\frac{e^{-\sqrt{\frac{s}{D}}|a|}}{2\sqrt{sD} (1+ \frac{K_0 S(0) \epsilon} {\sqrt{s D}})}.
\end{equation}
\noindent This when substituted back into Eq. (12) gives
\begin{equation}
\tilde P(x,s)=\frac{e^{-\sqrt{\frac{s}{D}}|x+a|}}{2\sqrt{sD}} -\frac{K_0S(0)\epsilon e^{-\sqrt{\frac{s}{D}}(|a|+|x|)}}{2\sqrt{sD}(\sqrt{s D}+K_0S(0) \epsilon)}.
\end{equation}
Now we have an explicit formula for $\tilde P(x,s)$, our interest is to calculate the survival probability $P(t) = \int_{-\infty}^{\infty} P(x,t)dx$. It is possible to calculate the Laplace transform of $\tilde P(s)$ of $P(t)$ directly. Again $\tilde P(s)$ is related to $\tilde P(x,s)$ by $\tilde P(s) = \int_{-\infty}^{\infty} \tilde P(x,s) dx$.
\begin{equation}
\tilde P(s)=\frac{1}{s} -\frac{\frac{K_0S(0)\epsilon}{\sqrt{D}} e^{-\sqrt{\frac{s}{D}}|a|}}{s(\sqrt{s} + \frac{K_0S(0) \epsilon}{\sqrt{D}})}.
\end{equation}
\noindent Using Inverse Laplace transformations \cite{Chen}, the expression for $P(x,t)$ is calculated as follows :
\begin{equation}
P(t)=1- e^{\frac{K_0S(0) \epsilon |a|}{D}+\frac{{{{K_0}^2} {{{S(0)}^2}} {\epsilon}^2} t}{D}}erfc\left({\frac{K_0S(0) \epsilon \sqrt{t}}{\sqrt{D}}}-{\frac{|a|}{2\sqrt{Dt}}}\right) +erfc(\frac{-|a|}{2\sqrt{Dt}}  
\end{equation}
So we have derived an analytical expression for survival probability in time domain for a particle diffusing under a flat potential with a localized sink. 

\section{Conclusions}

\noindent We have given a general analytically solvable model for calculating the survival probability as a function of time for a particle diffusing under a flat potential with a new localized sink. The exact analytical expression for survival probability has also been derived. The same method can be extended to more complex and realistic potential surfaces including a piece-wise linear potential \cite{HemCPL}  or a parabolic potential \cite{HemJCP} with the prerequisite that it's Green's function in absence of sink is known. 

\section{Acknowledgments}
\noindent One of the Authors would like to thank Ms. Moumita Ganguly and R. Sarvanan for their constant support and encouragement and DST for providing INSPIRE mentorship grant .

\end{document}